\documentstyle[aps,pre,multicol,epsfig]{revtex}
\begin{document}
\draft
\title{Analyzing symmetry breaking within a chaotic quantum system via
       Bayesian inference}
\author{C. I. Barbosa and H. L. Harney\\
        }
\address{Max-Planck-Institut f\"ur Kernphysik, D-69029 Heidelberg, 
         Germany\\        
        }
\date{\today}
\maketitle              
\begin{abstract}
Bayesian inference is applied to the level fluctuations of two coupled microwave
billiards in order to extract the coupling strength. The coupled resonators
provide a model of a chaotic quantum system containing two coupled symmetry 
classes of levels. The number variance is used to quantify the level
fluctuations as a function of the coupling and to construct the conditional
probability distribution of the data. The prior distribution of the coupling
parameter is obtained from an invariance argument on the entropy of the
posterior distribution.
\end{abstract} 
\begin{multicols}{2}
%\onecolumn
\narrowtext

\section{Introduction}
\label{I}

The subject of the present paper is Bayesian inference as applied to the
experiment of Ref.~\cite{PRL} in order to extract the mean square matrix 
element
coupling two chaotic classes of quantum states. The Bayesian procedure 
described
below does not contain any arbitrary element: 
The prior distribution --- sometimes left to
the educated guess of the analyst \cite{...} --- is determined by an invariance
argument on the entropy of the posterior distribution.

The present article is organized as follows. In Sec.~\ref{II}, we briefly 
describe
the experiment with superconducting microwave resonators that has 
provided the data
for the present analysis. The random matrix model for the coupling of two
symmetry classes of chaotic states is defined in Sec.~\ref{III}. It yields --- in
analytic form --- the dependence of the observable  on the coupling strength
which is to be determined. Bayesian inference, especially the definition of the
prior distribution, is discussed  in Sec.~\ref{IV}. The conditional 
probability
distribution of the data is defined in Sec.~\ref{V}. The results are given in 
Sec.~\ref{VI}. A discussion in Sec.~\ref{VII} concludes the paper.

\section{The experiment with coupled microwave resonators}
\label{II}

Billiards provide models of classical and quantum mechanical chaos. They have
been studied extensively, see the review article \cite{1}. Quantum mechanical 
billiards can be
simulated by flat microwave resonators \cite{2,3,4,5}. One class of these
``quantum" billiards are the Bunimovich stadium billiards 
\cite{buni} experimentally investigated in Refs.
\cite{PRL,5,D20,D23,D28,D29}. 

These investigations show that the fluctuation properties of the
quantum chaotic systems with well defined symmetries are described by Dyson's
matrix ensembles \cite{1}. In the case of the stadium billiards, the correct
description is provided by the Gaussian Orthogonal Ensemble (GOE). This means,
e.g., that the fluctuations of the positions of the eigenmodes --- shortly the
level fluctuations --- are the same as the fluctuations of the eigenvalues of
random matrices drawn from the GOE. In order to assess these fluctuations,
various statistics have been defined --- such as the distance of neighboring 
levels
or the variance of the number of levels in a given interval. The expectation 
values of these statistics have been worked out \cite{French,Leitner} 
for comparison with data such as the present ones.

\begin{figure} [hbt]
\centerline{\epsfig{figure=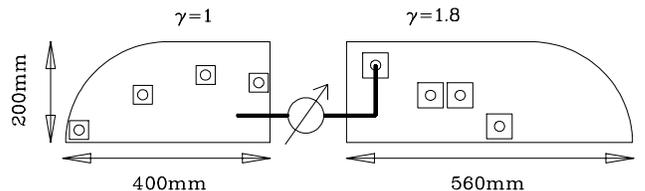,width=8.6cm,bbllx=2.2cm,bblly=18cm,
bburx=18.5cm,bbury=23.3cm}}
\caption{Shapes of the two coupled stadium billiards. The circles
inside the squares indicate the positions of the antennas used to scatter
microwave power through the system in order to find the eigenfrequencies of the
entire system. The parameter $\gamma $ is the ratio between the length of the
rectangular part and the radius of the circular part of the resonator. The
vertical heights of the stadia are given in Sec.~\ref{VI}.}
\label{logo}
\end{figure}

In the previous work \cite{PRL}, the level positions of a system 
have been measured that consisted of two (quarters of)
stadium billiards coupled electromagnetically. See Fig.~\ref{logo}. The
technical realization of the coupling has been described in Ref.~\cite{PRL}. 
In the
frequency range of 0 to 16 GHz, the complete spectra of the two stadia 
displayed
608 and 883 resonances in the $(\gamma=1)$ stadium and the
$(\gamma=1.8)$ stadium, respectively. The mean level spacing is $D=10.7$ MHz. 

\begin{figure} [t!]
\centerline{\epsfig{figure=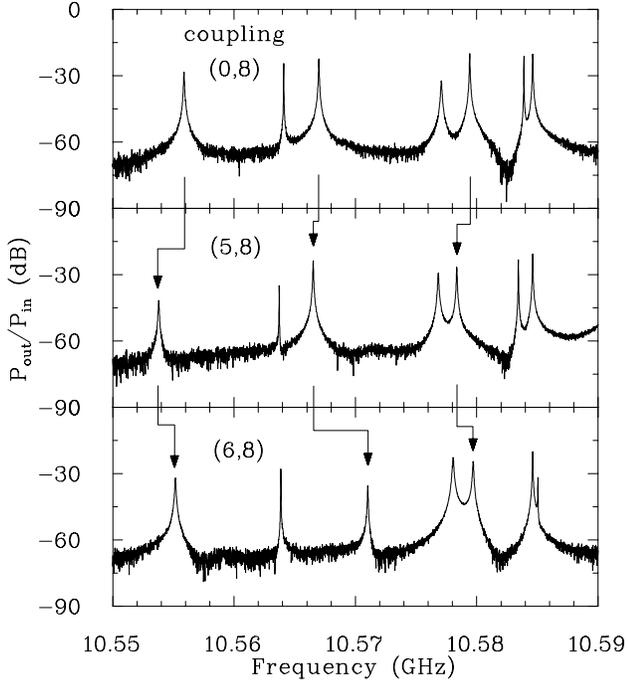,width=8.6cm,bbllx=3cm,bblly=3.5cm,
bburx=19.5cm,bbury=22cm}}
\caption{Three spectra --- within a small range of frequencies --- taken at
different couplings between the two resonators. The coupling increases from top
to bottom. Its notation $(x,y)$ is explained in Sec.~\ref{VI}. 
The arrows shall help to recognize the shifts of a few resonances.}
\label{Spektrum}
\end{figure}

In Fig.~\ref{Spektrum}, small pieces of spectra are shown for three different
couplings. The arrows shall help to recognize that --- due to the coupling ---
the resonances are shifted by statistically varying amounts.

This system simulates two symmetry classes of levels coupled by a symmetry
breaking interaction. Each class of levels --- represented by each of the
uncoupled stadia --- can be identified with a chaotic system of well defined
symmetry having the properties of the GOE. The entire system of the coupled
stadia no longer has the universal properties of the GOE. Its properties are 
a function of a suitably defined coupling parameter $\Lambda$.

The investigation of symmetry breaking in chaotic quantum systems is not a
recent challenge to physicists \cite{Ros-Porter}. Good examples of the
experimental and theoretical efforts already invested into this problem, are 
the
cases of isospin mixing \cite{9,10,11}, of parity violation in heavy nuclei
\cite{A6}, of the breaking of certain atomic and molecular symmetries
\cite{Ros-Porter,A8}. The experiment performed in \cite{PRL} provides a 
general model for
these case studies.

In the present paper, we do not describe any one of the specific case studies;
we shall not even describe in more detail the model experiment of 
Ref.~\cite{PRL}. We
rather describe --- in the next section --- 
the model experiment \cite{PRL} in an abstract mathematical form and then turn
to its analysis in Secs.~\ref{IV}-\ref{VII}.

\section{The mathematical model of symmetry breaking in a chaotic quantum 
system}
\label{III}

In the absence of coupling each eigenstate of the system of Fig.~\ref{logo} can
be characterized as belonging to either resonator 1 or resonator 2. This is
equivalent to the assignment of a quantum number $\gamma$. The spectrum of
states of each $\gamma$ has the statistical properties of the eigenvalues of
matrices drawn from the GOE. The superposition of the two spectra displays what
we shall call 2 GOE behavior. It can be described by a block-diagonal Hamilton
operator where each block is an element of the GOE, hence, by the first term of
the Hamiltonian
\begin {eqnarray}
  {\cal H}= \left( 
  \begin{array}{cc} 
     \framebox[0.9cm]{\rule[-0.5mm]{0cm} {0.1cm}$\mbox{GOE}$} & 0\\
     0 & \framebox[0.9cm]{\rule[-0.5mm]{0cm}{0.1cm}$\mbox{GOE}$}
  \end{array}\right) 
  +  \alpha \left( 
  \begin{array}{cc} 
     \framebox[0.5cm]{\rule[-0.1mm]{0cm} {0.1cm}0} &  V\\
     V^+  &   \framebox[0.5cm]{\rule[-0.1mm]{0cm}{0.1cm}0}
  \end{array}\right).  
\label{Block}
\end {eqnarray}
For $\alpha \neq 0$, the off-diagonal matrix $V$ in the second term on the
r.h.s.~provides the coupling between both symmetry classes. It has Gaussian
random elements --- as the GOE blocks. If the two GOE blocks have the same
dimension then their elements as well as the elements of $V$ shall all 
have the same rms
value. Then $\alpha =1$ turns ${\cal{H}}$ as a whole into a GOE matrix
\cite{footnote}. The resulting spectrum displays what 
we call 1 GOE behavior.
If the two GOE blocks have different dimensions, then the rms values must be
chosen such that their spectra have the same length. The details are given in
\cite{Leitner}. This model is a special case of the model by Rosenzweig and Porter
\cite{Ros-Porter}. 

The parameter that governs the level statistics is $\alpha v/D$ rather than 
$\alpha$. Here, $D$ is the mean level distance of $\cal H$. See Refs.
\cite{Leitner,11}. In the sequel, the coupling parameter 
\begin{eqnarray}
\Lambda = \left( \frac{\alpha v}{D} \right) ^{2}
\label{3}
\end{eqnarray} 
will be used. Often the coupling strength is also parametrized in terms of the
spreading width
\begin{eqnarray}
\Gamma ^{\downarrow } &=& 2 \pi \, \frac{(\alpha v)^{2}}{D} \nonumber \\
&=& 2\pi \, \Lambda D   .
\label{4}
\end{eqnarray}

The statistic used in the present paper in order 
to characterize the behavior of the data, is the so-called
$\Sigma ^{2}$ statistic or number variance. It is the variance $z(L)$ of the
number $n(L)$ of levels found in an interval of length $LD$, i.e.
\begin{eqnarray}
z(L) = \langle (n(L) - L)^{2} \rangle .
\label{5}
\end{eqnarray}
Here, the angular brackets $\langle \rangle$ denote the average over all pieces
of spectra of length $L$ that have been cut out of the entire experimental
spectrum. The procedure is described in Sec.~\ref{V}. 

The expectation value $\overline {z(L)}$ with respect to the statistical
ensemble defined by Eq.~(\ref{Block}) is called $\Sigma^{2}(L,
\Lambda  )$. This function has
been calculated by French et al.~\cite{French} and by Leitner et al.
\cite{Leitner}. According to \cite{Leitner}, it is
\begin{eqnarray}
\Sigma ^{2} (L,\Lambda )  &=& \overline{z(L)} \nonumber \\
&=& \Sigma ^{2} (L,\infty ) + \frac{1}{\pi ^{2}} \ln 
\left( 1 + \frac{\pi ^{2} L^{2}}{4(\tau + \pi ^{2} \Lambda )^{2}} \right) .
\label{6}
\end{eqnarray}
Here, $\Sigma ^{2} (L, \infty )$ is the expression
\begin{eqnarray}
\Sigma ^{2} (L, \infty)  &=& \frac{2}{\pi ^{2}} \left\{ \ln (2\pi L) 
+ \gamma _{E} +1
+ \frac{1}{2} \left[ {\mbox{Si}} (\pi L) \right] ^{2} \right. \nonumber \\
& & - \frac{\pi }{2} {\mbox{Si}} (\pi L) - \cos (2\pi L) - {\mbox{Ci}} (2\pi L)
\nonumber \\
& & + \left. \pi ^{2} L \left[ 1 - \frac{2}{\pi } {\mbox{Si}} (2\pi L)\right] 
\right\} .
\label{7}
\end{eqnarray}
It describes the 1 GOE behavior. The second term on the r.h.s.~of Eq.~(\ref{6})
obviously vanishes for $\Lambda \rightarrow \infty$.

In Eq.~(\ref{7}), $\gamma _{E}$ is Euler's constant and ${\mbox{Si}}$, ${\mbox{Ci}}$ are
the sine and cosine integrals defined e.g.~in paragraph 8.23 of \cite{GR}. The
parameter $\tau $ is a function  of the ratio between the dimensions of the two
GOE blocks in the first term of Eq.~(\ref{Block}). In the present situation, it is
equal to 0.74.

\begin{figure} [hbt]
\centerline{\epsfig{figure=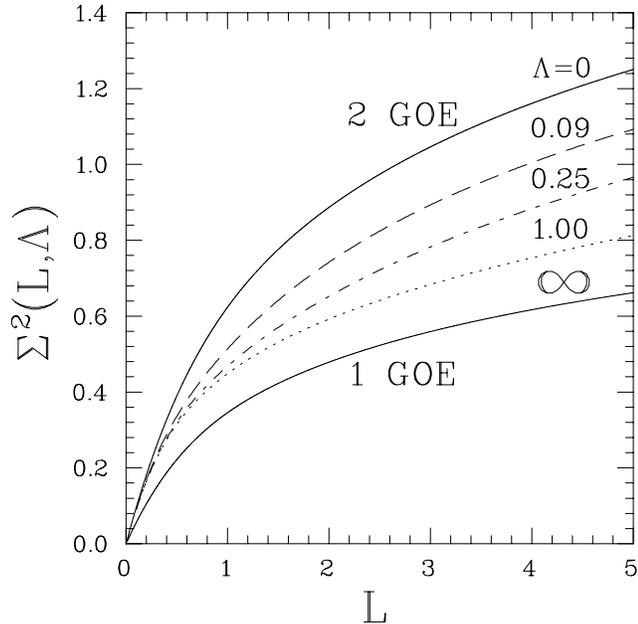,width=8.6cm,bbllx=2cm,bblly=2.3cm,
bburx=19cm,bbury=19.5cm}}
\caption{The expectation value $\Sigma ^{2} (L, \Lambda )$ of the number
variance $z(L)$ as a function of $L$ for various coupling strengths $\Lambda$
--- ranging from 2 GOE to 1 GOE behavior. The abscissa has been limited to 
$L \leq 5$ because the data analyzed below are in the range of $1 \leq L \leq 5$.}
\label{Sigma2bild}
\end{figure}

The function $\Sigma ^{2}$ depends on the coupling parameter $\Lambda $ --- as
is illustrated by Fig.~\ref{Sigma2bild}. Therefore $\Lambda $ can be inferred
from the experimental number variance $z(L)$ . The principle of this inference
is described in the next section.

\section{Bayesian inference}
\label{IV}

Suppose that a set of experimental data $z_{k}$, $k=1...M$, is given which
depends on a parameter $\Lambda $ in the sense that the probability distribution
$w_{k}$ of the event $z_{k}$ is conditioned by the hypothesis $\Lambda $,
\begin{eqnarray}
w_{k} = w_{k} (z_{k} \mid \Lambda) .
\label{8}
\end{eqnarray}
The events $z_{k}$ shall be statistically independent of each other. The joint
distribution $W$ of the $z_{k}$, $k=1...M$, is then
\begin{eqnarray}
W(z \mid \Lambda ) = \displaystyle \prod ^{M}_{k=1} w_{k} (z_{k} \mid \Lambda ).
\label{9}
\end{eqnarray}
From this follows the distribution $P(\Lambda \mid z )$ of $\Lambda $  under the
condition that the data $z$ are given via Bayes' theorem
\begin{eqnarray}
P(\Lambda \mid z) = \frac{W(z \mid \Lambda) \, \mu (\Lambda )}{m(z)} .
\label{10}
\end{eqnarray}
Here, $\mu(\Lambda )$ is the so-called prior distribution. It is the measure of
integration in the space of $\Lambda $. One must define it such that it
represents ignorance on $\Lambda $ --- in a sense described below.
The function $m(z)$ is the prior
distribution of $z$. It is not independent of $\mu $; it is given by the
normalizing integral
\begin{eqnarray}
m(z) = \int d\Lambda \,\,  W(z \mid \Lambda ) \, \mu (\Lambda ).
\label{11}
\end{eqnarray}
In the framework of the logic underlying Eq.~(\ref{10}), a probability
distribution of --- say --- $\Lambda $ is considered to represent the available
knowledge on $\Lambda $ and the prior distribution corresponds to ``ignorance
about $\Lambda $". 

The definition of $\mu (\Lambda)$ 
deserves a detailed comment. First of all, 
the natural choice of $\mu (\Lambda )$
is not the constant function because a reparametrization $\Lambda \rightarrow
\lambda (\Lambda )$ will transform $\mu (\Lambda )$ into\begin{eqnarray}
\displaystyle
\mu _{T} (\lambda ) = \mu (\Lambda ) \left| \frac{d\Lambda }{d\lambda} \right| .
\label{12A}
\end{eqnarray}
Unless the transformation is linear, it turns a uniform distribution into a
non-uniform one.

We define $\mu
(\Lambda )$ such that the entropy of $P(\Lambda | z )$
does not depend on the true value $\hat{\Lambda }$ that governs the 
distribution of the data $z= (z_{1}...z_{M})$. 
The data follow the distribution $W(z |\hat{\Lambda })$. Although 
$\hat{\Lambda }$ is not known, it is supposed to be a well defined number. If it 
is shifted to another value $\hat {\Lambda }^{\prime }$ and one takes new data
$z^{\prime }$ and constructs the posterior distribution 
$P^{\prime }=P(\Lambda | z ^{\prime })$ 
from the new data, then one can expect $P^{\prime }$ to be 
shifted with respect to $P$. The distribution $P^{\prime }$  will be 
centered in the vicinity of $\hat{\Lambda }^{\prime }$ rather than 
$\hat {\Lambda }$. However, we want to make sure that the ``spread" of 
$P^{\prime }$ is the same as that of $P$; that is, the entropy of 
$P$ and $P^{\prime }$ shall be the same --- 
for a given number of data $M$. 
In this sense, no value of $\Lambda $ is a priori preferred over 
any other one.

The definition of the entropy requires some attention. The usual formula
$- \int d\Lambda \, P \ln \, P$ for the entropy
is of too restricted validity in the present context, because this 
expression is not
invariant under a reparametrization $\Lambda \rightarrow \lambda
(\Lambda )$. The general expression for the entropy is
\begin{eqnarray}
\displaystyle
H= - \int d\Lambda \, P(\Lambda | z) \, \ln \frac{P(\Lambda | z)}{\mu (\Lambda
)}
\label{13A}
\end{eqnarray}
which is independent of a reparametrization \cite{JAY63,JAY68}, because the
transformations of both distributions, $P$ and $\mu $, are performed according
to (\ref{12A}). Therefore the derivative $| d\Lambda / d \lambda |$
drops out of the argument of the logarithm and expression (\ref{13A}) is left
unchanged by the substitution $\Lambda \rightarrow \lambda $.

It is possible to define $\mu $ such that $H$ is independent of the true 
value $\hat {\Lambda }$, if $W$ possesses the property introduced in
\cite{JAY68,HART,STEIN} called form invariance. It states that there
is a group of transformations $\cal{G}_{\rho}$ such that the simultaneous
transformation of $z$ and $\Lambda $ leaves $W$ invariant, i.e.
\begin{eqnarray}
\displaystyle 
W({\cal{G}}_{\rho} z | {\cal{G}}_{\rho} \Lambda ) \, d{\cal{G}}_{\rho} z = 
W(z | \Lambda )\, dz .
\label{14A}
\end{eqnarray}
The group parameter $\rho $ must have the same domain of definition as the
hypothesis $\Lambda $. If one chooses $\mu (\Lambda )$ to be the invariant
measure of the group then it is not difficult to show that the posterior
distribution $P$ also possesses the invariance (\ref{14A}). This entails that
$H$ is invariant under any transformation $z \rightarrow {\cal{G}}_{\rho} z $ of
the data. However, by Eq.~(\ref{14A}) this is just what happens to a given data
set if the true value $\hat{\Lambda }$ is shifted to $\hat {\Lambda }^{\prime} =
{\cal{G}}_{\rho}^{-1} \hat{\Lambda }$. 

There is a handy formula that yields the invariant measure without any study of
the structure of the group. It is
\begin{eqnarray}
\displaystyle 
\mu (\Lambda ) = \left| M^{-1} 
\int d^{M}z \, W(z \mid \Lambda ) \frac
{\partial ^{2}} {\partial \Lambda ^{2}} \ln W(z \mid \Lambda ) \right| ^{1/2} 
\label{15A}
\end{eqnarray}
and was proposed by Jeffreys \cite{FIS} even before form invariance was
discussed.

Not every conditional distribution $W$ possesses a symmetry (\ref{14A}). Even if
this is not the case, expression (\ref{15A}) ensures that $H$ 
is approximately independent of the true value of $\Lambda $. This holds in the
following sense: For every $\hat{\Lambda }$ one can replace the correct
distribution $W(z \mid \Lambda )$ by an approximation $W_{app}
(z \mid \Lambda )$ which is form invariant. The approximate and the correct
distributions agree within the fourth order of $\Lambda - \hat{\Lambda }$.
Equation (\ref{15A}) yields the invariant measure of the approximation to within
the third order of $\Lambda - \hat{\Lambda }$ \cite{Harney}.

In summary: expression (\ref{15A}) ensures that no value of $\Lambda $ is a
priori preferred over any other one if the form invariance (\ref{14A}) exists. 
If there is no form invariance,
expression (\ref{15A}) approximately ensures this. 
Therefore (\ref{15A}) is the best recommendation in any
case.

Neither the group theoretic argument nor Jeffreys' rule nor information
theoretic arguments are new in the discussion of the Bayesian prior. However,
the way in which they are related justifies the present digression on a
fundamental issue. We omit to show how and why the present arguments are related
to the geometric considerations which were introduced by Amari \cite{AMA} and
are currently put forward by Rodriguez \cite{ROD}. These authors agree on the
result (\ref{15A}).

The posterior distribution $P$ is used to construct an interval of error 
often called a confidence interval. It is the shortest interval that
contains $\Lambda $ with probability $K$. The usual error is defined with the
confidence $K=0.68$.

The posterior distribution $P$ approaches a Gaussian for $M \rightarrow \infty $
provided that the true value of $\Lambda $ is not on the border of the domain of
definition of $\Lambda $. One can prove that 
the variance of the Gaussian is proportional to
$M^{-1}$. Hence, with increasing $M$ the posterior distribution $P$ 
will become so narrow that $\mu $ changes very little
in the domain where $P$ is essentially different from
zero. Note that $\mu $ does not depend on $M$. 
Then $\mu $ drops out of expression (\ref{10}). If this happens, the
present Bayesian analysis becomes equivalent to a $\chi ^{2}$ fit of $\Sigma
^{2} (L,\Lambda )$ to the experimental points $z(L)$. The standard procedure of
the $\chi ^{2}$ fit can e.g.~be found in \cite{numrec}. It does not require a
prior distribution.

If $P$ is not Gaussian, the $\chi ^{2}$ fit yields meaningless confidence
intervals. Then Bayesian  inference cannot be bypassed. In the example presented
below this happens in the limit of small coupling between the resonators:
Eventually, the posterior distribution $P$ decreases monotonically. The
experiment is then compatible with zero coupling because the shortest confidence
interval contains the point $\Lambda =0$ for any $K$. 
The point of zero coupling is on the border of the domain of definition of
$\Lambda $.

\section{The distribution of the data}
\label{V}

Spectral fluctuation properties can only be studied after secular variations of
the level density have been removed, i.e.~after the frequency scale has been
transformed such that the level density becomes unity within the interval
covered by the experiment. This procedure --- often called ``unfolding" the
spectrum --- is a standard one \cite{UNF} and has been applied.

After this we defined --- for a given interval of length $L$ --- 
$N_{L}$ adjacent intervals. The
intervals did not overlap and no space was left in between them. This means

\begin{eqnarray}
N_{L} = \left[ \frac{{\mbox{length of spectrum}}}{L} \right],
\label{16}
\end{eqnarray}
where the square brackets designate the largest integer contained in the
fraction. For each interval, the number $n(L)$ of levels occuring within it was
counted and the squared difference $(n(L) - L)^{2}$ was averaged over the
$N_{L}$ intervals. This defines the average $\langle \rangle$ introduced in 
Eq.~(\ref{5}) and, 
hence, the ``event" $z(L)$. This procedure was repeated for a set
of values $L_{k}$, $k=1...M$, to be defined below. In this way, $M$ events 
\begin{eqnarray}
z_{k} \equiv z(L_{k})
\label{17}
\end{eqnarray}
were obtained.

The Bayesian procedure outlined in the previous section requires that one
assigns a probability distribution $w_{k} (z_{k} \mid \Lambda )$ to each event.
The $z_{k}$ are statistical quantities in the following sense: If another
spectrum would be provided that had the same statistical properties as the
measured one and the data  $z_{k}$ would be constructed in the same way as
above, they would of course not coincide with the data obtained from the 
actually
measured spectrum --- precisely because the levels are subject to statistical
fluctuations. If one could go through the ensemble of spectra in this way, one
would obtain an ensemble of data $z_{k}$. We are looking for the distribution
$w_{k}$ of this ensemble. Since there is only the single measured spectrum and
since no theory yielding $w_{k}$ is available, we have generated the 
distribution of $z_{k}$ by Efron's bootstrap method \cite{boot}. This method
generates the distribution numerically by drawing 
at random and independently a new set of $N_{L_{k}}$ intervals from the
$N_{L_{k}}$ original intervals. A new $z_{k}$ is produced from this new set of
intervals.
Repeating this many times, a distribution of
$z_{k}$ is generated which is identified with the distribution $w_{k}$ of the
$z_{k}$. Note that $N_{L_{k}}$ is always large, namely $N_{L_{k}} \gtrsim 300$. 

For $L_{k} \geq 1$, the distribution $w_{k}$ was in this way found to be a
$\chi ^{2}$ distribution with $N_{L_{k}}$ degrees of freedom --- which
intuitively seems reasonable. They are close to Gaussians with variance
$2/N_{L_{k}}$
As mentioned above in Sec.~\ref{III}, the mean value of
this distribution is 
\begin{eqnarray}
\displaystyle \overline{z_{k}} &=& \int dz \,\, z \, w_{k} (z \mid \Lambda ) 
\nonumber \\
\displaystyle &=& \Sigma ^{2} (L_{k}, \Lambda ) .
\label{18}
\end{eqnarray}
Since $\Sigma ^{2}$ depends on $\Lambda $ --- see Eq.~(\ref{6}) --- the
distribution $w_{k}$ depends on $\Lambda $.

We have restricted the analysis to $L_{k} \geq 1$. In the domain of 
$L_{k} <1$, the number variance so weakly depends on the parameter $\Lambda $
that one does not give away much information by this restriction.

\begin{figure} [hbt]
\centerline{\epsfig{figure=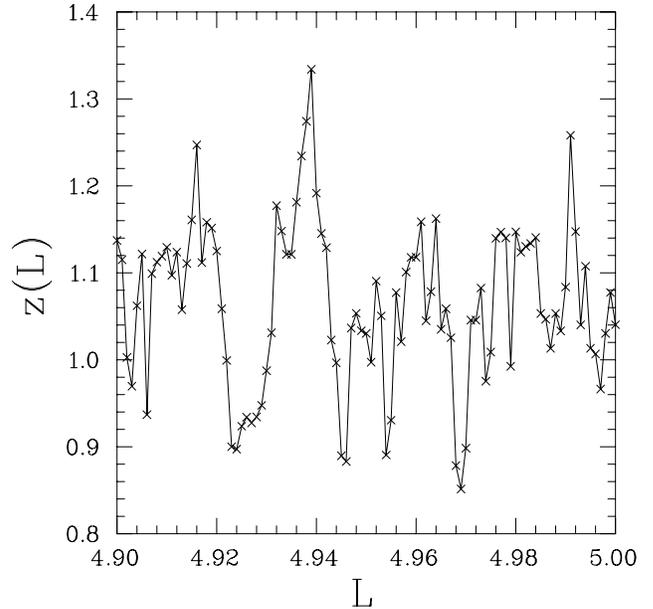,width=8.6cm,bbllx=2cm,bblly=2.3cm,
bburx=19.3cm,bbury=19.4cm}}
\caption{The experimental estimate $z(L)$ of the number variance --- see 
Eq.~(\ref{5}) --- as a function of $L$ calculated in steps of 
$\Delta L = 0.001$.
The typical width of the structures in this fluctuating function determines the
range $\epsilon$ over which $z(L)$ and $z(L+\epsilon)$ may be correlated.
Conversely, it serves to estimate the minimum distance $\epsilon $ between $L$
and $L^{\prime}$ which must be respected if $z(L)$ shall not be correlated with
$z(L^{\prime})$.}
\label{step}
\end{figure}

In order to avoid an unnecessarily complicated 
distribution of the $z_{k}$, we want to be sure that there are no 
correlations between $z_{k}$
and $z_{k ^{\prime}}$, for $k \neq k ^{\prime}$. It was therefore necessary to
determine the minimum $\epsilon $ of the distance $\mid L_{k} - L_{k ^{\prime}}
\mid $ that would still allow for statistically independent $z_{k}$, 
$z_{k ^{\prime}}$. Indeed if $\mid L_{k} - L_{k ^{\prime}}
\mid $ is very small then most of the intervals associated with $L_{k}$ will
almost coincide with an interval associated with $L_{k^{\prime}}$. As a
consequence, many of the numbers $n(L_{k})$ found in the intervals associated
with $L_{k}$ will occur also in the intervals associated with $L_{k^{\prime}}$.
Hence, $z_{k}$ will not be independent from $z_{k ^{\prime}}$. In order to
determine $\epsilon $, we have calculated $z(L)$ as a function of $L$ in steps
of 0.001. For a small range of $L$, the result is given in Fig.~\ref{step}.
Indeed over a distance of a few times this step width, $z(L)$ changes little. If
$\mid L - L^{\prime} \mid $ is many times this step width, then $z(L)$ and
$z(L^{\prime})$ show independent fluctuations. In principle, one can study the
decay of the correlations as a function of $\mid L - L^{\prime} \mid $ by
constructing the autocorrelation function of $z(L)$. We have contented ourselves
to inspect Fig.~\ref{step} and similar plots for different domains of $L$. It
seems obvious from Fig.~\ref{step} that the typical width of the structures is
less than 0.025. This justifies to set
\begin{eqnarray}
\epsilon = 0.025,
\label{19}
\end{eqnarray}
to define
\begin{eqnarray}
L_{k} = 1+ (k-1) \epsilon , \,\,\,\,\,\,\, k=1,2,... M \, \, ,
\label{20}
\end{eqnarray}
and to assume that $z_{k}$  is statistically independent of $z_{k^{\prime}}$ for
$k \neq k^{\prime}$. 

There is an upper limit $L_{max}$ of $L$ that one must be aware of: The spectral
fluctuations of levels from billiards agree with those of random matrices ---
i.e.~they are universal --- within intervals of a maximum length which is
inversely proportional to the length of the shortest periodic orbit in the
billiard \cite{B85,DEL}. This requires $L_{max} =5$ here. 

Hence, data $z(L)$ for $1 \leq L \leq 5$ were used to obtain the results
presented below. This means that by Eqs.~(\ref{19},\ref{20}) the number of
statistically independent data points is $M=161$.

Let us note that one can devise definitions of the set of intervals with given
length $L$ other than adjacent
intervals --- as was done here.
One can admit a certain overlap between them as suggested in \cite{Bohigas-bins}
or one can place them at random \cite{HOFF}. We have tried these alternatives
and have made sure that they do not significantly change the results presented
below.

\section{Results}
\label{VI}

\begin{figure} [hbt]
\centerline{\epsfig{figure=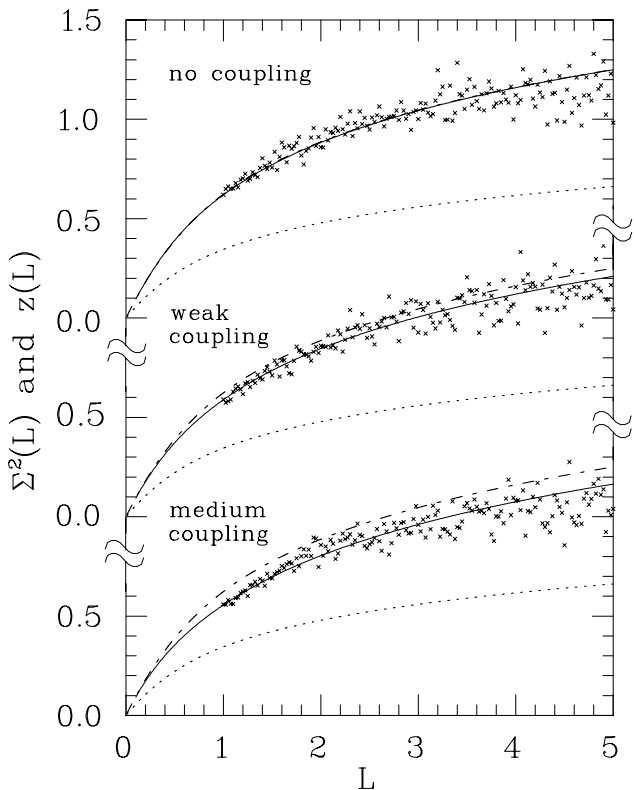,width=8.6cm,bbllx=1.3cm,bblly=0.5cm,
bburx=19cm,bbury=25.5cm}}
\caption{The number variance $z(L)$ (crosses) and its expectation value $\Sigma
^{2} (L,\Lambda )$ (full lines) for various experimental couplings $(x_{1},
x_{2})$. The dashed lines display the 2 GOE behavior, i.e.~$\Sigma ^{2} (L,0)$.
The dotted lines give the 1 GOE behavior, i.e.~$\Sigma ^{2} (L,\infty )$.}
\label{coupling}
\end{figure}

The data $z(L_{k})$, $k=1,...,161$, are given on Fig.~\ref{coupling} for the six
different couplings that were experimentally investigated. The coupling
strength increases from top to bottom on Fig.~\ref{coupling}. Its experimental
realization is indicated by the two numbers $(x_{1}, x_{2})$ in brackets that
label the six parts of the figure. They are explained as follows: 
The billiards were positioned with their flat sides against each other.
Holes were drilled through the walls of the
resonators such that a niobium pin could be inserted
perpendicularly to the plane of the billiards through the $(\gamma =1.8
)$ stadium into the $(\gamma =1 )$ stadium. The coupling strength is
determined by the depths $x_{1}$ and $x_{2}$ by which the niobium pin penetrates
into the $(\gamma =1)$ and the $(\gamma =1.8)$ stadium, respectively. These
depths are given by $(x_{1},x_{2})$ in mm. The net height of the ($\gamma
=1$) stadium was 7 mm and that of the $(\gamma =1.8)$ stadium was 8 mm. 
For the strongest coupling --- i.e.~the bottom part of
the figure --- a second niobium pin, penetrating all the way through both
resonators, was added. The coupling (0,8) --- i.e.~the top part of the figure
--- is the case, where the billiards should be decoupled. 

The dashed lines on Fig.~\ref{coupling} illustrate the limiting case of 2 GOE
behavior, i.e.~expression (\ref{6}) with $\Lambda =0$. The dotted lines show
the limit of 1 GOE behavior, i.e.~expression (\ref{7}). Obviously, all six
cases are not easily distinguished from the 2 GOE behavior, i.e.~$\Lambda =0$.

Prior to the analysis it was therefore not clear whether the six experimental
cases would yield distinguishable coupling parameters $\Lambda $ and whether
these would even be distinguishable from zero. The latter question means 
according to Sec.~\ref{IV}:
It was not clear whether a $\chi ^{2}$ fit would be appropriate.
Therefore the whole analysis was based on Bayesian inference. 
The prior distribution was calculated from (\ref{15A}). The probability 
distribution $w_{k} (z_{k} | \Lambda )$ of the data has been defined in
Sec.~\ref{V}. Whether form invariance exists has not been investigated.
The scatter of the data is quite large ---
especially for $L$ close to 5. These fluctuations are assessed by the
distribution $w_{k} (z_{k} \mid \Lambda )$. The
fluctuations increase with increasing $L$. This is reflected by the fact that
$w_{k}$ was found to be a $\chi ^{2}$ distribution with $N_{L_{k}}$ degrees of
freedom. Its relative rms deviation is $\sqrt{2/N_{L_{k}}}$ and $N_{L_{k}}$
decreases with increasing $L_{k}$, see Eq.~(\ref{16}). Despite the scatter of
the data the coupling parameter is so well determined that the analysis
distinguishes the six experimental cases from each other --- because the
number of data points is large enough.

For all cases except coupling (0,8), the posterior distribution (\ref{10})
turned out to be Gaussian. This is
illustrated on Fig.~\ref{(5,3)} for the coupling (5,3). 
In the case of coupling (0,8) --- which is expected to show 2 GOE behavior ---
the posterior distribution  of $\Lambda $ is the monotonically decreasing
function of Fig.~\ref{(0,8)}. This is reasonable because the shortest confidence
interval on $\Lambda $ will --- for any confidence --- include the possibility
of $\Lambda =0 $. Hence, the distribution of Fig.~\ref{(0,8)} allows to
state only an upper limit for $\Lambda $. 

\begin{figure} [b!]
\centerline{\epsfig{figure=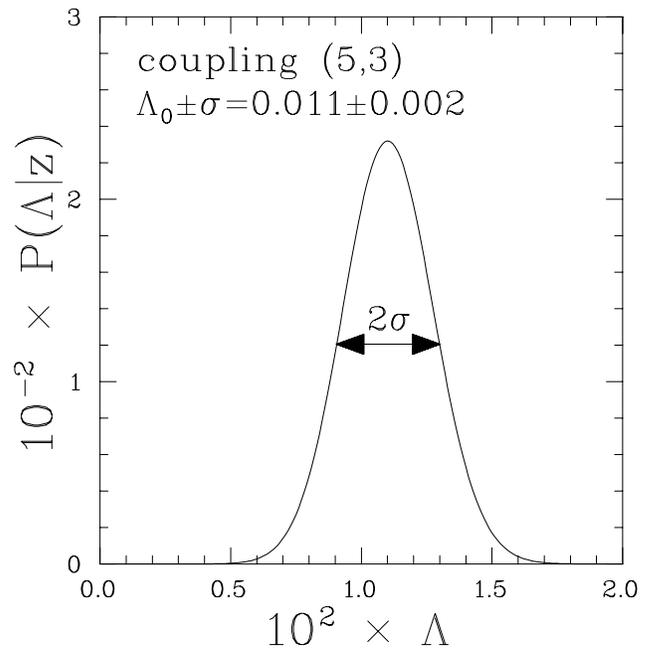,width=8.6cm,bbllx=2.5cm,bblly=2.3cm,
bburx=19cm,bbury=19.3cm}}
\caption{Posterior distribution for the coupling (5,3), i.e.~the second case on
Fig.~\ref{coupling} and in Table \ref{alpha_NND}. The center $\Lambda _{0}$ 
and the rms deviation of the Gaussian are specified.}
\label{(5,3)}
\end{figure}

\end{multicols}
\onecolumn
\widetext
\begin{table}
\caption {Parameters found for six different experimental couplings. The results
          have been obtained via Bayesian inference as outlined in 
	  Sec.~\ref{IV}. The column headings are explained in the text. 
	  \label{alpha_NND}}
\begin{tabular}{ c | cccc | cc }
   Exp.~coupling & $\Lambda $ & $\Gamma^\downarrow / D$ & $\alpha v/D$ & 
   $\alpha v \,\,\,\,\,\,\,\,\,\,\,\,$ & 
   $\chi ^{2}$ & $\Lambda _{\mbox{fit}}$  \\
\tableline
  (0,8) & $\le 0.00085$ & $ \le 0.005$ & $\le 0.029$ &  
  $\le 0.31\,\,\,\,\,\,\,\,\,\,\,\,$ & 
  $1.14$ & -  \\
  (5,3) & $0.011\pm0.002$ & $0.07\pm0.01$ & $0.105\pm0.008$ & 
  $1.12\pm0.09\,\,\,\,\,\,\,\,\,\,\,\,$ & 
  $0.90$ & $0.013\pm0.002$ \\
  (4,4) & $0.017\pm0.002$ & $0.11\pm0.01$ & $0.130\pm0.007$ & 
  $1.39\pm0.07\,\,\,\,\,\,\,\,\,\,\,\,$ & 
  $1.04$  & $0.019\pm0.002$ \\
  (5,8) & $0.030\pm0.002$ & $0.19\pm0.01$ & $0.173\pm0.006$ & 
  $1.85\pm0.06\,\,\,\,\,\,\,\,\,\,\,\,$ & 
  $1.11$  & $0.033\pm0.002$ \\
  (6,8) & $0.032\pm0.002$ & $0.20\pm0.01$ & $0.180\pm0.006$ & 
  $1.93\pm0.06\,\,\,\,\,\,\,\,\,\,\,\,$ & 
  $1.47$  & $0.037\pm0.003$ \\
  (6,8)+(7,8) & $0.040\pm0.002$ & $0.25\pm0.01$ & $0.200\pm0.006$ & 
  $2.14\pm0.06\,\,\,\,\,\,\,\,\,\,\,\,$ & 
  $1.24$ & $0.044\pm0.002$ \\
\end{tabular}
\end{table}
\vspace*{-0.5cm}
\begin{multicols}{2}
\narrowtext

The results of the Bayesian analysis are summarized in the first five columns of
Table I. The first column characterizes the
experimental realization of the coupling as explained above. In the
second column, the coupling parameter $\Lambda $ is given. It can also
be expressed (in the third column) by the ratio $\Gamma ^{\downarrow}/D$,
see Eq.~(\ref{4}). Alternatively --- see Eq.~(\ref{3}) --- the combination
$\alpha v/D$ of parameters of the model of Eq.~(\ref{Block}) is given in
the fourth column. By putting $D$ equal to the mean level distance
$D=10.7$ MHz of the experiment, one obtains in the fifth column the rms
coupling matrix element $\alpha v$ in MHz. 

In the case of coupling (0,8), where only an upper limit for the coupling can be
given, we have done so --- for the confidence of 68\%. In all other cases the
center $\Lambda _{0}$ of the Gaussian posterior is given together with the rms
deviation; this defines a 68\% confidence interval.

\begin{figure} [hbt]
\centerline{\epsfig{figure=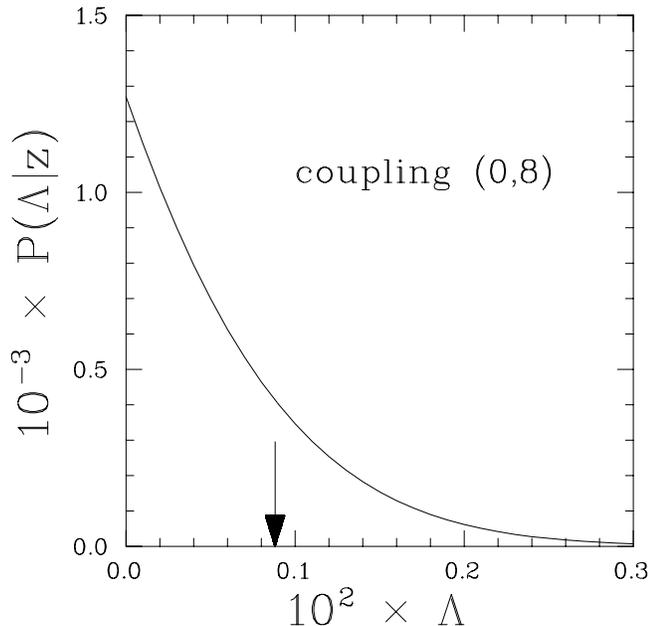,width=8.6cm,bbllx=2cm,bblly=2.2cm,
bburx=19cm,bbury=19cm}}
\caption{Posterior distribution for the coupling (0,8), i.e.~the case where the
coupling is expected to be zero. The probability integrated from $\Lambda = 0 $
to 0.00085 (marked by an arrow) is 68\% .}
\label{(0,8)}
\end{figure}

According to Sec.~\ref{IV}, Gaussian posteriors suggest that one may replace
Bayesian inference by a $\chi^{2}$ fit which is simpler. A $\chi ^{2}$ fit has
been performed in all cases and the results are given in the last two columns of
the table. The sixth column displays the normalized $\chi ^{2}$ value.
For a reasonable fit, it should lie between 0.84 and 1.16. 
This follows from the fact that the distribution of $\chi ^{2}$ is here
approximately Gaussian with rms value $(2/(\mbox{number of degrees of
freedom}))^{1/2} = (2/160)^{1/2} \approx 0.16$. The seventh 
column gives the coupling parameters $\Lambda $ which the fit has found. They
are compatible with the Bayesian results except for the first entry (0,8). Here
the fit puts out a negative value, i.e.~it does not produce a meaningful result.
This was expected from the discussion in Sec.~\ref{IV}.

\section{Discussion}
\label{VII}

The emphasis of the present paper is on the Bayesian
analysis of the data. Although Bayes' theorem provides a clear and simple
prescription of how to draw conclusions from data 
about a hypothesis conditioning the
data, its use was hampered for a long time by the difficulty to define the prior
distribution $\mu $ of the hypothesis. Equation (\ref{15A}) is a very general
definition of $\mu $. It applies even to cases, where the variable $z$ of the
events is discrete (the integral in (\ref{15A}) then means a sum). The prior
distribution (\ref{15A}) ensures that the amount of information one gets on the
hypothesis $\Lambda $ is --- at least approximately --- 
independent of the true value of $\Lambda $. 

Supplemented by Eq.~(\ref{15A}), Bayes' theorem provides the generalization of
all methods of inference that rely on Gaussian approximations. The method of the
least squares e.g.~belongs to them. It does not require a prior distribution of
the parameter to be determined. In the present paper the relation
between Bayesian inference and $\chi ^{2}$ fit has been
discussed. A criterion has been given under which Bayesian inference is
approximately equivalent to the simpler fit procedure. This criterion has been
substantiated numerically.

The present formalism especially provides the correct treatment of
null-experiments, i.e.~of experiments that yield only an upper limit for the
parameter of interest. An example for this situation has been presented.
By the same token, the formalism of Sec.~\ref{IV} provides the decision whether
the parameter is compatible with zero.

The physical results of the present analysis show that the strongest coupling
realized in the microwave experiment \cite{PRL} has about the same size as the
coupling found in \cite{11} to occur between states of different isospin in
$^{26}$Al. The strongest coupling treated in the present paper causes about 25\%
mixing between the two classes of levels, i.e.~a state which can be
approximately assigned to the ($\gamma=1$) stadium contains about 25\% strength
from the configurations of the ($\gamma =1.8$) stadium --- and vice versa. This
is the interpretation of the value of $\Gamma ^{\downarrow}/D$ in Table I. Data
that are as numerous and precise as those of Ref.~\cite{PRL} allow to
detect $\Gamma ^{\downarrow}/D$ ten times smaller than the result of \cite{11} 
--- according to the
present analysis. Nuclear data ---  which never provide as large a sample of
states as the experiment \cite{PRL} --- would not allow to detect 
$\Gamma ^{\downarrow}/D = 0.07$ (the smallest detected mixing in Table I) from
the level fluctuations. The precision obtained in this experiment has allowed 
to detect the subtle
dependence of the level fluctuations on the breaking of a symmetry which is
predicted by the random matrix model \cite{11,French,Leitner}.

\acknowledgements

The authors thank Dr. T. Guhr for helpful discussions. They thank Prof. 
H. A. Weidenm\"uller for his support and advice.They are indebted to
Prof. A. Richter and the members of the ``chaos group" of the Institut f\"ur
Kernphysik at Darmstadt, H. Alt, H.-D. Gr\"af, R. Hofferbert, and H. Rehfeld, 
for their help and encouragement. 
One of the authors (C.I.B.) acknowledges the financial support granted 
by the Fritz Thyssen Stiftung and the CNPq (Brazil).

\end{multicols}


\begin{references}
\bibitem {footnote} Note that there is a lack of precision in the text of 
                      \cite{PRL} below its Eq.~(1). The condition of equal 
		      dimensions of the GOE blocks is missing.
\end{references}
\end{document}